\documentclass[
twocolumn,
showpacs,superscriptaddress,amssymb,aps,prl]{revtex4}

\usepackage{graphicx}
\usepackage{bm}

\begin{document}
\title{Oscillatory Magneto-Thermopower and Resonant Phonon Drag \\in a
High-Mobility 2D Electron Gas}
\author{Jian Zhang}
\affiliation{Department of Physics, University of Utah, Salt Lake City, Utah 84112}
\author{S. K. Lyo}
\affiliation{Sandia National Laboratories, Albuquerque, New Mexico 87185}
\author{R. R. Du}
\affiliation{Department of Physics, University of Utah, Salt Lake City, Utah 84112}
\author{J. A. Simmons}
\affiliation{Sandia National Laboratories, Albuquerque, New Mexico 87185}
\author{J. L. Reno}
\affiliation{Sandia National Laboratories, Albuquerque, New Mexico 87185}

\date{October,10, 2003}
\begin{abstract}
Experimental and theoretical evidence is presented for new low-magnetic-field
($B<5$ kG) $1/B$-oscillations in the thermoelectric power of a high-mobility
GaAs/AlGaAs two-dimensional (2D) electron gas. The oscillations result from
inter-Landau-Level resonances of acoustic phonons carrying a momentum
equal to twice the Fermi wavenumber at $B = 0$. Numerical calculations show
that both 3D and 2D phonons can contribute to this effect.               
\end{abstract}
\maketitle

Thermoelectric power (TEP) of a two-dimensional electron gas (2DEG) reflects
the electron density of states, scattering dynamics, and
electron-phonon interactions.  Phonon scattering and consequently the TEP
are particularly important in GaAs/Al$_x$Ga$_{1-x}$As heterostructures owing 
to the strong electron-lattice coupling in these materials. The TEP
experiments in the following two limits of the magnetic field ($B$) have been
widely pursued
\cite{fletcher4,nicholas,davidson,fletcher3,fletcher2,ying,zeitler,bayot,fletcher16,btieke}
and the phenomena in these regimes are relatively well-understood
\cite{cantrell,lyo38,kuba,lyo40,smith}.
  At $B = 0$, the TEP ($S_0$) shows a power-law, e.g., $T
^{3-4}$, temperature dependence above $T\sim0.3$ K, indicating that the
phonon drag
mechanism dominates and the electron-diffusion effect is relatively weak.
  In the regime of a high field ($B>10$ kG),
Shubnikov de-Haas (SdH) oscillation and quantum Hall effect have been
observed in TEP \cite{btieke, fletcher16}. The peak
amplitude of the TEP are found to be greater than the                       
predicted values of the diffusion TEP by 2 orders of
magnitude, but consistent with the phonon-drag TEP. Most
of the above experimental data at $B=0$ and at relatively high fields have been
successfully explained quantitatively by theories based on the phonon drag
TEP.

On the contrary, little is known experimentally about the TEP in the
regime of a weak magnetic field, where many Landau levels (LLs)
are occupied by electrons with large quantum numbers $n\gg1$ at the Fermi
level, and electronic transport is generally treated semiclassically. This
regime is characteristically distinct for the following reasons. 1)
The Fermi wave length is much shorter than the magnetic length
$l_B=\sqrt{\hbar/eB}$ (i.e., $2k_F\ell_B\gg1$), and a momentum selection rule
governs the scattering of electron guiding centers. In particular, $2k_F$
scattering is strongly enhanced, giving rise to the low
$B$ resonance phenomena                                               
\cite{zudovlip,yang} in this regime.  2) Acoustic phonons in GaAs
have suitable energies, and in combination with 1) can participate in
inter-LL scattering. This is dramatically different from
the regime of higher $B$, where intra-LL scattering dominates
and inter-LL scattering is negligible at low temperatures\cite{lyo40}.
Elastic intra-LL scattering is important for LL broadening.

In this Letter, we report low-$B$ TEP oscillations observed
in a high-mobility 2DEG in GaAs/Al$_{x}$Ga$_{1-x}$As
heterostructures. The oscillations are periodic in $1/B$ with the
peak positions of $B$ proportional to $\sqrt{n_e}$, or to the
Fermi wavenumber $k_F=
\sqrt {2\pi n_e}$, where $n_e$ is the electron density. Characteristically,
such oscillations appear in the temperature range 0.5 K - 1 K, and their
amplitude increases with $T$. It will be shown that the TEP oscillations
result from inter-LL cyclotron resonance promoted by resonant absorption of
phonons carrying a $2k_F$ in-plane momentum and an energy
$\ell\hbar \omega_c$ equal                               
to the integer ($\ell=1,2,\cdots$) multiple of $\hbar\omega_c=\hbar
eB/m^*$, which is the cyclotron energy
with an effective mass $m^*$. 2D phonon modes propagating along the
GaAs/Al$_x$Ga$_{1-x}$As interface and 3D phonons are responsible
for the TEP oscillations. However,  the contribution from 2D phonons is
difficult to assess quantitatively because of the lack of electron-2D phonon
interaction parameters. On the other hand, numerical
calculations show that 3D phonons  yield a substantial contribution to the
oscillation and behave like 2D phonons for the
$2k_F$-oscillation because $q_z$ is restricted to small
$q_z\ll2k_F$ at low temperatures due to the phonon occupation factor.

The TEP experiments were performed using a sorption pumped $^3$He cryostat
and a superconducting solenoid, with the B axis always oriented
perpendicular to the 2DEG plane. The samples used here are made from
GaAs/Al$_{0.3}$Ga$_{0.7}$As
heterostructures grown by molecular beam epitaxy on the (001) GaAs
substrate. At low temperature ($T\sim1$K), the $n_e$ and the mobility,   
$\mu$, can be varied using a red light-emitting diode (LED). Without LED
(saturate LED) the $n_e$ (in unit of $10^{11}$cm$^{-2}$ throughout this
paper) $\sim$ 1.33 (2.03) and $\mu$ $\sim$ 2 $\times 10^6$ (3 $\times
10^6$) cm$^2$/Vs. To make TEP measurement, a specimen of dimension 8 mm
$\times$ 2 mm was first cleaved from the wafer and a Hall bar mesa (width
0.5 mm) was then chemically wet-etched from it by optical lithography. Heat sink
to a copper post (0.3 K) was achieved by indium soldering at one end of
the specimen.
A strain gauge was glued to the
other end of the specimen by Ag paint and was used as a heater to create a
temperature gradient ($\nabla T$) along the Hall bar direction. Two
calibrated RuO$_2$
chip sensors were glued by epoxy 2850FT \cite{2850} on the back of
the specimen and used to
measure the $\nabla T$ along the sample. Electrical leads were made
with 38 gauge manganin wires, whose low thermal conductance ensures a
negligible heat leak to the $^{3}$He liquid. The whole system was sealed
in a
vacuum can made of epoxy \cite{1266}. The vacuum can including the copper
cold sink were immersed in the $^{3}$He liquid.

The TEP $S_{xx}$ is defined by $\nabla V_{xx}=S_{xx}\nabla T$, with
the quantities measured in the following manner.
A low frequency ($f_{0}$ = 2.7Hz) ac voltage was applied to the heater and
the $T_{1}$, $T_{2}$ (see inset, Fig. 1) were measured using the
RuO$_{2}$ sensors and a AC bridge. The $V_{xx}$ induced by the
thermo-gradient was measured by lock-in method at the frequency of
$2f_{0}$ = 5.4 Hz. Both the $T$ and voltage gradients were
calculated using the dimensions given by the specimen.

\begin{figure}
\includegraphics{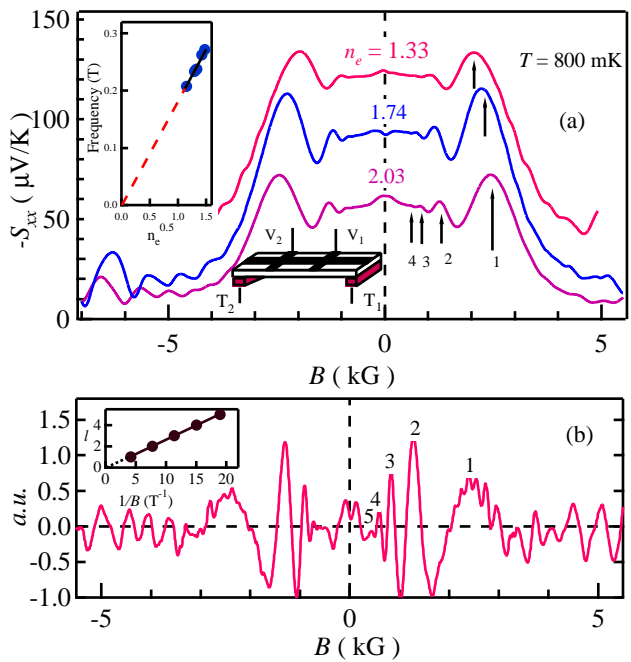}
\caption{
-$S_{xx}$ traces are shown for three densities $n_{e}$ of 1.33, 1.74,    
and 2.03 in units of 10$^{11}$cm$^{-2}$, respectively; arrows indicate the
maxima
for $l$ =1, 2, 3, 4 and the shift of the primary ($l$ = 1) peak with
increasing $n_{e}$.
In the lower figure, the second derivative against $B$ for the
high-density
trace is shown; the numbers mark the oscillation peaks. Inset shows that
oscillations are periodic in $1/B$.}
\label{fig1}
\end{figure}

In Fig.\,1(a), we show the low field magneto-TEP measured at 800
mK, for
three electron densities. These traces
reveal strong new oscillations appearing at $B < 3$ kG, where SdH
oscillations are relatively weak. For example, up to four maxima can be 
clearly seen in the trace of $n_e$ = 2.03. The arrows close to the
traces indicate the maxima (indexed as $l$ = 1, 2, 3, 4). A second
derivative with respect to $B$, - $d^{2}S_{xx}/ dB^{2}$, for the $n_e$ =
2.03 is plotted in (b), together with a fan diagram (inset) showing the
linear relation between the order, $l$, and the inverse $B$. We conclude
from all three traces that the TEP oscillations are periodical in
$1/B$. Moreover, the peak positions of $B$ scales with $\sqrt{n_e}$, as
shown in the inset of
Fig.\,\ref{fig1}(a).   This behavior is distinct from that of SdH which scales
with
$n_e$, but consistent with the characteristics of a class
\cite{zudovlip,yang} of weak $B$ oscillations originating from cyclotron
resonance with a $2k_F$ momentum transfer.

\begin{figure}
\includegraphics{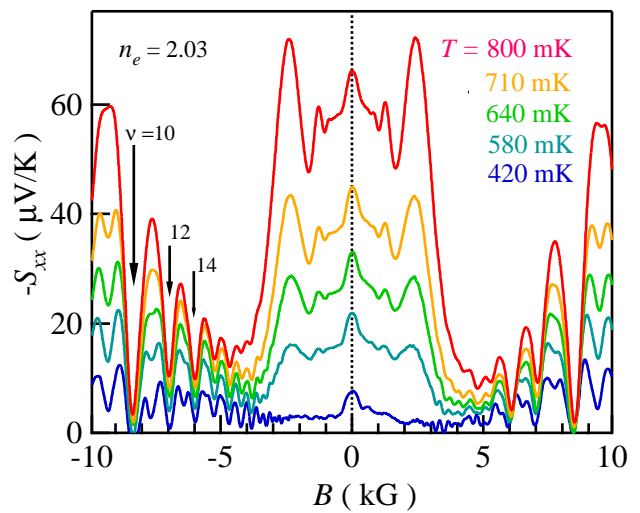}
\caption{\label{fig2}
-$S_{xx}$ (for $n_{e}$ = 2.03$\times$10$^{11}$cm$^{- 2}$) traces at 
different temperatures show that the low field oscillations become
stronger as $T$ increases.
}
\end{figure}

The TEP oscillations exhibit a remarkable $T$ dependence, which,
as will be shown later, can be attributed to inter-LL scattering by
acoustic phonons. As an example, Fig. 2 shows the TEP data for $n_e =
2.03$ at a $T$ range from 420 mK to 800 mK. Note that the TEP oscillation
can be discerned at $T$ as low as 300 mK for our experiment. With increasing
$T$, the TEP signal increases dramatically, both at $B$ = 0, and at $B <
5$ kG. The TEP at $B = 0$, $S_0$, is well understood as due to phonon
drag. As will be shown, the $S_0$ is power-law dependent in this
experiment. As $T$ rises, the TEP oscillation amplitude increases faster
than $S_0$. This fact indicates that inter-LL scattering is dominant in
the TEP oscillations observed here, since inter-LL scattering mechanism
leads to an exponential rather than a power-law $T$                    
dependence. As  the cyclotron energy  increases, the inter-LL
scattering regime with an exponential-law $T$-dependence disappears and only
the intra-LL-scattering SdH regime with a power-law
$T$-dependence prevails\cite{lyo40}. Indeed, at higher
temperatures,
$T\ge 800$ mK, the TEP oscillations become the dominating feature in
TEP measurement at low field.

\begin{figure}
\includegraphics{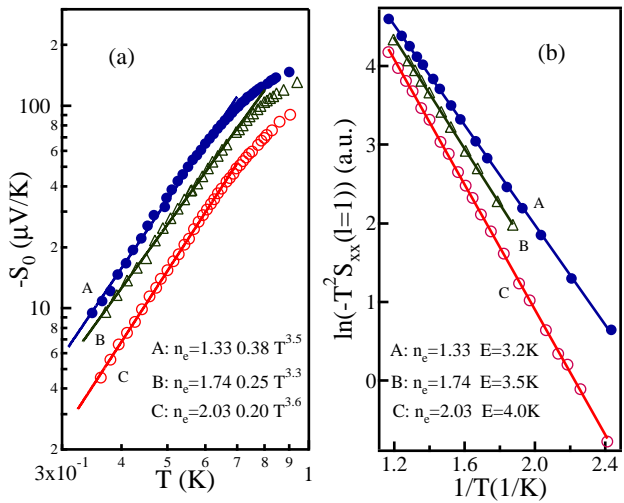}
\caption{\label{fig3}
The left figure shows the $T$ dependence of the TEP at zero 
field ($S_{0}$), for
three densities. The solid lines represent $T^3$ temperature
dependence. The right figure shows the $T$ dependence of
the TEP at primary oscillation maxima ($l=1$), for the three
corresponding densities. The E is the fitted value for the phonon energy
at $l=1$.}
\end{figure}

We now turn to a quantitative analysis for  $S_0$ and the TEP
oscillations. In
Fig.\,\ref{fig3}(a), we plot $S_0$ vs. $T$ for three
densities. All data show a power-law dependence on $T$, with an exponent
between 3 and 4. This observation confirms quantitatively  that the TEP at
$B$ = 0 is by the phonon drag mechanism \cite{btieke}. The
$T$-dependent amplitude of the first peak ($l = 1$) is presented in
Fig.\,\ref{fig3}(b). It is worth noting that the data strongly deviates from
a power-law, but can be fitted by a modified exponential relation
$S_{xx}\sim exp(-E/k_B T)/T^2$, predicted by equation
(1). From the slope of the fit we arrive at the activation energies $E \sim$
3.2 K, 3.5 K, 4.0 K, respectively for the three densities, which are
somewhat smaller than $\hbar\omega_c\sim$ 4.0 K, 4.4 K, 4.8 K due to the LL
broadening. These data strongly support the interpretation of a resonant
phonon drag mechanism being the origin of the TEP oscillations.

In principle, both 3D and 2D phonons can contribute to the resonant
phonon drag mechanism responsible for the oscillations. We start by
considering the 3D case, in the following, on a more general 
ground. Qualitatively, the 2D case can be reduced from the 3D case.

We begin with the formula for the phonon drag TEP in a magnetic
field derived by Lyo \cite{lyo40}; a similar formulism was given by
Kubakaddi
and Butcher \cite{cantrell}. The TEP is obtained for a unit volume:

$$S_{xx}={{-k_Bh} \over {e\nu(k_BT)^2}}
\sum_{s{\bf q}}\sum_{n,n^\prime} u_s\Lambda_{s{\bf q}}q_y^2\ n_{s{\bf 
q}}|V_{s{\bf
q}}|^2 \Delta_z(q_z)\Delta_{n,\ell}(q_\|)$$
\begin{equation}\times\int d\varepsilon\rho_n(\varepsilon)\int
d\varepsilon^\prime\rho_{n^\prime}(\varepsilon^\prime)
f(\varepsilon)[1-f(\varepsilon^\prime)]\delta (\varepsilon+\hbar
\omega _{s{\bf q}}-\varepsilon^\prime),\label{e1}
\end{equation}
where $\nu=\pi n_el_B^2$ is the filling factor for spin-degenerate LLs,
$\Lambda_{s{\bf q}}$ the phonon
mean-free-path, $n_{s{\bf q}}$   the boson function,
$f(\varepsilon)$ the Fermi function, $u_s$ the sound velocity for the mode
$s$, $\rho_n$ the spectral function for the LLs, and $\Delta_z(q_z)$ is the
conservation factor for $q_z$\cite{lyo40}.  The square of the absolute value
of the electron-phonon matrix element
$|V_{s{\bf q}}|^2$ is proportional
to the
in-plane-momentum conservation factor:
\begin{equation}
\Delta_{n,\ell}(q_\|)=\frac{n!}{(n+\ell)!}\chi^\ell
e^{-\chi}[L_n^\ell(\chi)]^2,\ \ \chi=\frac{(q_\|l_B)^2}{2},
\label{e2}
\end{equation}
which has a sharp
principal maximum near $\chi=4n$, namely, near the in-plane momentum
transfer $q_\|\simeq 2k_{\rm F}$ for
$n \gg 1,\ \ell $ in view of $\varepsilon_F\simeq
n\hbar\omega_c$ \cite{zudovlip,holstein}. In Eq.\,(\ref{e2}),
$L_n^\ell(\chi)$ is the associated Laguerre polynomial and
$n^\prime=n+\ell$ is the larger of $n$ and $n^\prime$. The phonon occupation
factor  restricts
$q_z$ to small values
$q_z\ll 2k_{\rm F}$ for this resonance at low temperatures. There
are other secondary peaks below the main peak for $\Delta_{n,\ell}(q_\|)$. To
reduce the computing time, we approximate the spectral density function
$\rho_n(\varepsilon)$  by a rectangular distribution with a full width
$2\Gamma$ centered  at  the LL energy
$\varepsilon_n=\hbar\omega_c(n+1/2)$ and take $\Gamma
= 0.2$ meV for numerical evaluation. 

The calculated TEP is plotted as a function of $B$ in Fig. 4, for
$\Lambda_{s{\bf q}}=2$ mm and respectively for three densities, 2.03,
1.74, and 1.33$\times 10^{11}$ cm$^{-2}$ employing  field-free screening for
the electron-phonon interaction. Other parameters are well-known and are given
in Ref.\cite{lyo40}.  The TEP is proportional to the phonon mean-free-path
$\Lambda_{s{\bf q}}$, which is basically an adjustable parameter. At low
temperatures considered here,
$\Lambda_{s{\bf q}}$ is determined by boundary scattering and is expected to
be of the order of the smallest sample dimension $\sim 2$\ mm. It is seen
that the
$\ell=1,2,3,\cdots$ peaks appear on top of the regular SdH
oscillations which are from  the phonons with $q_\|< 2k_F$.
Larger
$\Gamma$ makes the peaks broader and the valleys shallower compared to the
sharper structures obtained for
$\rho_n(\varepsilon)=\delta(\varepsilon-\varepsilon_n)$.  A, B and C traces
display the TEP as a function of
$B$ for three densities
$n_{e}=1.33,\ 1.74,\ 2.03\times 10^{11}$ cm$^{-2}$ at $0.8$ K.  It is seen 
that there is an approximate scaling relationship between the peak
positions of $B$, satisfying
$B\propto \sqrt{n_{e}}/\ell$. This
relationship is consistent with the inter-LL resonance phonon picture
$\hbar\omega_{2k_F}\simeq\ell\hbar eB/m^\ast$ in view of
$\omega_{2k_F}\propto k_{\rm F}\propto\sqrt{n_{e}}$, yielding reasonable
agreement with experimental data. We also find $S_{xx}\propto
\exp(-E/k_BT)/T^2$ in agreement with Fig.\,\ref{fig3}(b) with
$E\propto\sqrt{n_e}$ close to the transverse $2k_F$ phonon energy. Transverse
phonons yield a dominant contribution ($\sim 70$ \%) through strong
piezoelectric scattering at low temperatures. The calculated background TEP
is much lower than the peaks compared with the data in Fig.\,\ref{fig2}
probably due to the simplistic non-self-consistent density of states employed
in the present low-$B$ situation where the LLs are closely separated. Also,
the magnitude of the calculated TEP keeps decreasing as
$B$ approaches
$B=0$ in contrast to the data: Below $B<0.4$ kG, the number of the LLs
becomes very large ($n>100$), requiring a zero-$B$ formalism for a more
accurate result.  

\begin{figure}
\includegraphics{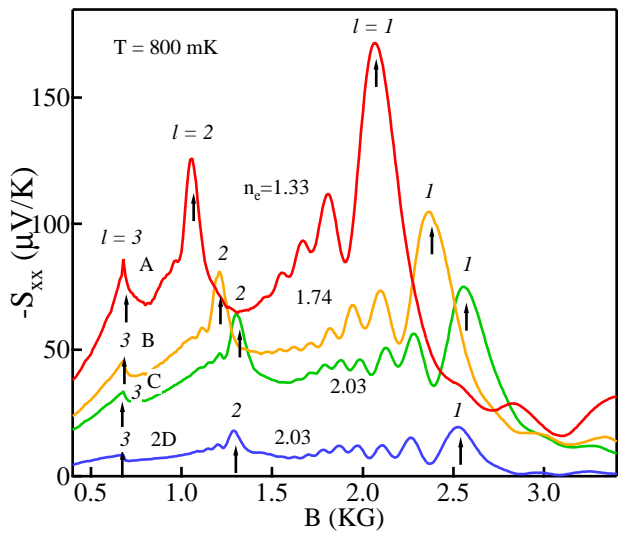}
\caption{\label{fig4}
Numerical calculation of the TEP at low fields based on the 3D phonon
model (upper three curves). The bottom curve is an estimate from a
2D-phonon model. The primary oscillation maxima increase with
$T$, decrease with increasing $n_e$. Their $B$ positions shift to lower field  
with decreasing $n_e$.}
\end{figure}

The 2D phonon modes relevant in the GaAs/Al$_x$Ga$_{1-x}$As
heterostructures are the leaky phonons \cite{zudovlip,efros} with  phonon
wave vector component $q_{z}$ = 0 and $q=q_\|$. The TEP is again given by
Eq.\,(\ref{e1}) with $\Delta(q_z)\equiv 1$ with the summation on $q_z$
replaced by the summation over the leaky modes.  For a rough estimate, we
take same $|V_{s{\bf q}}|$ with the effective sample volume given by
$\Omega=S\ell_p$, where $S$ is the cross section of the well and
$\ell_p$ is roughly the penetration depth of the mode. The result for
the 2D phonons is compared with that of the 3D phonons in Fig.\,\ref{fig4} for
$\ell_p=200$ $\AA$ using a pair of longitudinal and transverse modes. The $B$
dependence of $S_{xx}$ from the 2D phonons in Fig.\,\ref{fig4} is very similar
to that from the 3D phonons except that it is slightly shifted to lower $B$.

In Fig.\,\ref{fig3}(a), the linear contribution $S_0\propto T$ from
the electron-diffusion TEP is negligible at $B=0$. This situation is
similar to the data of Tieke, $et$ $al$.
\cite{btieke}, but different from the results of C. Ruf, $et$ $al$. \cite{ruf}, in which the
diffusive TEP is visible at 0.6K and the data deviate from the $T^3$ slope at  
0.6K.

A high-mobility sample is necessary to observe this low-field
acoustic-phonon resonance. In high mobility samples, high LLs
are not easily smeared out by impurity-scattering. We expect stronger and
sharper $2k_F$ oscillations for higher-mobility samples.

In conclusion, we have reported for  the first time an oscillatory
TEP in a weak magnetic field, where inter-LL
scattering is accessible by acoustic phonons. The observation of such
oscillations confirms a generic  $2k_F$ momentum selection rule
in electronic transport in a weak magnetic field where many LLs
are occupied. Finally, it is found that while both 3D and 2D phonons cause
qualitatively similar TEP oscillations in this regime, 3D phonons yield a
substantial contribution to the oscillations to explain the data.

We acknowledge helpful conversations with Prof. R. Fletcher.  S. K. L thanks
Dr. D. E. Amos for his indispensable help with the computation.   The work at
the University of Utah was supported by NSF and by a DARPA-QuIST
grant. Sandia is a multiprogram
laboratory operated by Sandia Corporation, a Lockheed Martin Company, for 
the U.S. DOE under Contract No.DE-AC04-94AL85000.

\bibliographystyle{unsrt}

\end{document}